\documentstyle[preprint,aps]{revtex}

\title{Kaon flow as a probe of the kaon potential in nuclear medium} \bigskip
\author{G. Q. Li, C. M. Ko, and Bao-An Li}
\address{\it {\ Cyclotron Institute and Physics Department,\\
Texas A\&M University, College Station, Texas 77843}}

\begin{document}
\maketitle

\begin{abstract}
The flow of kaons, i.e., the average kaon
transverse momentum as a function of rapidity,
and the associated flow parameter in heavy-ion collisions at SIS/GSI
energies is investigated in the relativistic transport model. It is found
that the pattern of kaon flow at the final stage is sensitive to the kaon
potential used in the model and is thus a useful probe of the kaon
potential in a nuclear medium.
\end{abstract}

Kaon properties in dense nuclear matter is a subject of considerable current
interest. This knowledge is useful for our understanding of both chiral
symmetry restoration in dense matter and the properties of the dense nuclear
matter existing in neutron stars \cite{BRO88,THO94} and formed in heavy-ion
collisions \cite{KO91,FANG93}. The possibility of $s$-wave kaon condensation
in dense nuclear matter was first predicted by Kaplan and Nelson \cite{KL86}
based on the mean-field approximation to the chiral Lagrangian. Since then,
there have been
numerous studies on the properties of kaons in nuclear matter, based on
various theoretical approaches such as the effective chiral Lagrangian \cite
{BRO87,PW91,MU92,BRO92}, the Nambu$-$Jona-Lasinio (NJL) model \cite{WEI92},
and the phenomenological off-shell meson-nucleon interaction \cite{YA93}.
The predictions from these models, however, have been inconsistent with each
other, especially concerning the higher-order corrections beyond the
mean-field approximation.

In the mean-field approximation to the effective SU(3)$_L\times $SU(3)$_R$
chiral Lagrangian \cite{KL86,PW91}, the kaon energy in a nuclear matter
satisfies
\begin{equation}
\label{disp}\omega ^2({\bf k},\rho )=m_K^2+{\bf k}^2-{\frac{\Sigma _{KN}}{%
f_K^2}}\rho _S+{\frac 34}{\frac \rho {f_K^2}}\omega ({\bf k},\rho ),
\end{equation}
where $m_K$ and {\bf k} are the kaon mass (in free space) and momentum,
respectively. The third term in the above equation is from the attractive
scalar interaction due to chiral symmetry breaking and is determined by
the kaon-nucleon sigma term $\Sigma _{KN}\approx 350$ MeV, the nuclear scalar
density $\rho _S$, and the kaon decay constant $f_K$, which is normally
assumed to be the same as the pion decay width, i.e., $f_K\sim
f_\pi \sim 93$ MeV. The
last term is from the repulsive vector interaction and is proportional to the
nuclear density $\rho $. From Eq. (\ref{disp}), we obtain the kaon energy in
a nuclear medium
\begin{eqnarray}\label{disp1}
\omega({\bf k}, \rho )=\Big[m^2_K+{\bf k}^2-{\Sigma _{KN}\over f_K^2}
\rho _S +\big({3\over 8}{\rho \over f_K^2}\big)^2\Big]^{1/2}
+{3\over 8}{\rho \over f^2_K}.
\end{eqnarray}

For low densities, including higher-order corrections beyond
the mean-field approximation would lead to a kaon
dispersion relation which is similar to that determined from the kaon-nucleon
scattering length using the impulse approximation \cite{rho94}, i.e.,
\begin{eqnarray}\label{disp2}
\omega({\bf k}, \rho )=\Big[m_K^2+{\bf k}^2-4\pi \big(1+{m_K\over m_N}\big)
{\bar a}_{KN} \rho \Big]^{1/2},
\end{eqnarray}
where $m_N$ is the nucleon mass and ${\bar a}_{KN}\approx -0.255$ fm is the
isospin-averaged kaon-nucleon scattering length in free space \cite{BS94}.

However, theoretical calculations \cite{BRO88A,CHEN92} indicate that the
simple impulse approximation (Eq. (\ref{disp2})) underestimates experimental
kaon-nucleus scattering data by 20-30\%. A more repulsive kaon-nucleon
interaction is needed to bring the theoretical results in better agreement
with the experimental data \cite{CHEN92}. In the Nambu$-$Jona-Lasinio model,
higher-order corrections are found to cancel out almost completely the
attractive scalar potential \cite{WEI92}, so the kaon feels effectively only
the repulsive vector potential. In this case, the kaon energy is
\begin{eqnarray}\label{disp3}
\omega({\bf k},\rho)
=\Big[m^2_K+{\bf k}^2+\big({3\over 8}{\rho \over f_K^2}\big)^2\Big]^{1/2}
+{3\over 8}{\rho \over f^2_K}.
\end{eqnarray}

In heavy-ion collisions, the dense matter formed is highly excited.
Higher-order corrections are thus expected to be suppressed, and the
mean-field approximation may be more appropriate. Indeed, for kaon
production in heavy-ion collisions at energies below its production
threshold from the nucleon-nucleon interaction in free space, it has been
found \cite{FA94} using the relativistic transport model \cite{KO87} that an
attractive kaon scalar potential is needed to explain quantitatively recent
experimental data by the Kaos collaboration \cite{GSI} from the heavy-ion
synchrotron (SIS) at GSI. Neglecting the kaon scalar attraction
would underestimate the kaon yield by about a factor of three.

Following Ref. \cite{SHU92}, we define the kaon potential in a nuclear
medium by
\begin{eqnarray}
U({\bf k}, \rho )=\omega ({\bf k}, \rho )-\omega _0({\bf k}),
\end{eqnarray}
where $\omega_0=\big(m_K^2+{\bf k}^2\big)^{1/2}$. The three potentials given by
Eqs. (\ref{disp1}), (\ref{disp2}), (\ref{disp3}) are all repulsive, as can
be seen in Fig. 1 where the density dependence of the kaon potential at zero
momentum is shown. The repulsion is relatively weak in the mean-field
approximation (solid curve)
because the repulsive vector potential is largely canceled by
the attractive scalar potential. The kaon potential is strongly repulsive,
if the scalar attraction is entirely canceled by higher-order corrections
(dashed curve) as
predicted in Ref. \cite{WEI92}. The result of the impulse approximation
using the kaon-nucleon scattering length (dotted curve)
lies between these two extreme
cases, which represent the lower and upper limits of current theoretical
predictions for the kaon potential in a nuclear medium.

To probe the kaon potential in a dense matter, further experimental
information will be very useful. In this Letter, we will show that the kaon
flow in central heavy-ion collisions is sensitive to the kaon potential, and
thus provides complementary, probably even more reliable, information on the
kaon potential in a dense matter. Particle flows in heavy-ion collisions have
been extensively studied in the past both experimentally \cite{GUT89,DEM90}
and theoretically \cite{DAN85,DAN93,BASS93,LI94}. The nucleon flow in both
symmetric and asymmetric heavy-ion collisions has been used to determine
separately the effects from the density and the momentum dependence of the
nucleon potential \cite{DAN93}. Pion flow has also been investigated in the
transport model \cite{BASS93,LI94}, and it has been found to be in the same
direction as the nucleon flow in central collisions but in the opposite
direction in peripheral collisions as a result of the strong absorption of
pions by nucleons. Recently, antiproton flow in heavy-ion collisions has
been studied using the Relativistic Quantum Molecular Dynamics \cite{GRE94}.
Because of the large annihilation cross section for antiprotons, the
antiproton flow is found to be opposite to that of nucleons.
Unlike pions and antiprotons,
kaons are not seriously affected by stochastic
two-body collisions as a result of their large mean free path.
The flow of kaons is thus a relatively clean probe for the kaon
potential in a nuclear medium.

Our study is based on the relativistic transport model developed in Ref.
\cite{KO87}. At incident energies around 1 GeV/nucleon as considered here,
the colliding system consists mainly of nucleons, deltas and pions. While
pions are treated as free particles, nucleons and deltas are propagated in
their mean-field potentials generated by the attractive scalar
and vector potentials as determined by
the non-linear $\sigma$-$\omega$ model \cite{QHD}. In the present work, we
use the so-called soft equation-of-state with a nucleon effective mass of
0.83$m_N$ and a compressibility of 200 MeV \cite{LI94B}.
The Schr\"odinger-equivalent potential of this model is in fair
agreement with that of the Dirac phenomenology when the nucleon kinetic energy
is below 800 MeV.  At higher energies, the explicit momentum dependence of
the scalar and vector potentials is needed, and the parameterization recently
proposed by the Giessen group would be more appropriate \cite{MARU94}.

Both elastic and inelastic reactions among nucleons, deltas and pions are
included using the standard Cugnon parameterization \cite{CUG82} and the
proper detailed-balance prescription \cite{DAN91}.
Because of its small production probability, kaon production in
heavy-ion collisions at subthreshold energies is treated perturbatively so
that the collision dynamics is not affected by the presence of produced
kaons \cite{AK85}. The kaon production cross section and its momentum
distribution from a baryon-baryon collision is taken from Ref.
\cite{RK81}. To improve the statistics for kaons, we have used
the perturbative test particle method of Ref. \cite{FA93}.

Representing kaons by test particles, their motions are given by the
following equations
\begin{eqnarray}
{d{\bf x}\over dt}={{\bf k}\over E^*}, ~~~~{d{\bf k}\over dt}=
-\nabla _x U ({\bf k}, \rho ).
\end{eqnarray}
In the above,
$U ({\bf k}, \rho )$ is defined in Eq. (5) with $\omega ({\bf k}, \rho )$
given by Eq. (\ref{disp1}), (\ref{disp2}), or (\ref
{disp3}), and corresponding $E^*$ given by $\omega ({\bf k}, \rho )
-(3/8)(\rho /f^2_K)$, $\omega ({\bf k}, \rho )$, or $\omega ({\bf k}, \rho
)-(3/8)(\rho /f^2_K)$,
depending on the assumptions made for the kaon potential.

The kaon-nucleon collision is included by using a kaon-nucleon total cross
section of about 10 mb \cite{DO82} which we take to be density-independent.
After the collision the kaon direction is isotropically distributed as the
kaon-nucleon interaction is mainly in the $s $-wave. Since kaon production
is treated perturbatively, its effect on nucleon dynamics is neglected. The
collision between a kaon and a pion via the $K^*$ resonance \cite{KO81} has
been neglected, as it is insignificant in heavy-ion collisions at 1
GeV/nucleon \cite{FA94}. We have also included the Coulomb interactions of
kaons with nucleons and pions.

We consider Au+Au collisions at an incident energy of 1 GeV/nucleon. In
order to suppress the rescattering effects from the spectators, only the
central collision at an impact parameter $b=$ 3 fm will be studied.
Our result for the nucleon flow
is in reasonable agreement with preliminary data from
the EOS collaboration \cite{EOS93} and with
those from calculations based on the normal Vlasov-Uehling-Uhlenbeck
model with a soft and momentum-dependent mean-field potential
\cite{DAN93}.

In Fig.
2 we show the average transverse momentum of kaons as a function of their
center-of-mass rapidity $y_{cm}$ at the final stage of the collisions. The
open circles, corresponding to the case without the kaon potential, show
that kaons flow in the same direction as nucleons, but with a smaller flow
velocity. The results using a strong repulsive kaon potential, corresponding
to Eq. (\ref{disp3}), are shown by solid circles.
The kaon flow in this case is in the
opposite direction from that of nucleons, i.e., the appearance of an
`antiflow' of kaons with respect to nucleons.
For a medium repulsion, as predicted by the impulse approximation
(Eq. (\ref{disp2})), the kaon flow shown by the open squares in the
midrapidity region is still opposite to the nucleon flow, but the `antiflow'
phenomenon is weakened. With a weak repulsion due to
both the scalar and vector interactions, we find that the
kaon flow shown by solid squares is in the same direction as that of the
nucleons, but follows the nucleons less closely than in the case without the
kaon potential. It is thus clear that the repulsive kaon potential tends to
make kaons flow away from nucleons. How large the separation of kaons from
nucleons is depends sensitively on the strength of the kaon potential. It is
therefore possible to study the kaon potential in a nuclear medium by
measuring the kaon flow in heavy-ion collisions.

To characterize the kaon flow more quantitatively, we introduce the
flow parameter $F$ defined by the slope of $<p_x>$ at midrapidity, i.e.,
$F={d<p_x>/dy_{cm}}|_{y_{cm}=0}$.
The time evolution of the kaon flow parameter is shown in Fig. 3.
Also shown in the figure by the
dashed curve is the time evolution of the central baryon density $\rho /\rho
_0$. In the early stage of the collision, there are very few kaons and a
reliable determination of the flow parameter is not possible. We have
therefore set it to zero until 8 fm/c. Kaons are mainly produced during the
time interval from 5 fm/c to 15 fm/c when the system is compressed and
baryon-baryon collisions are most energetic. During this period, kaons
follow essentially the flow of nucleons, so the flow parameter is positive
and similar to each other for the four cases considered here. The flow
parameter reaches its maximum value around 15 fm/c when the compression
almost ends. Afterwards the production of kaons is negligible and further
changes of the flow parameter are mainly due to the kaon potential and
kaon-nucleon scattering. In the expansion stage, clear differences are
observed among the kaon flow parameters corresponding to different
kaon potentials. Without the kaon potential, the final
kaon flow parameter ($\approx $ 52 MeV) is slightly smaller than its maximum
value as a result of kaon-nucleon scattering which are effectively
repulsive and tend to slightly repel kaons away from nucleons. Including
both scalar and vector potentials, the resulting kaon
potential is weakly
repulsive and the final flow parameter is still positive but is about a factor
of 3 smaller than the case without the kaon potential ($\approx 17 $ MeV).
Using the kaon potential based on the impulse approximation which is more
repulsive than the second case, the final flow parameter is negative and
small ($\approx -$9 MeV); i.e, in the mid-rapidity region kaons flow in the
opposite direction from the nucleons. Finally, if the scalar potential is
fully canceled by higher-order corrections, which leads to a strong
repulsive kaon potential, the kaon flow is negative and large ($\approx -$86
MeV). We would like to point out that the change of the kaon flow parameter
occurs mainly in the expansion stage from 15 fm/c to 22 fm/c, when the
central density is still above the normal nuclear matter density.  Since
the average density in the vicinity of kaons changes from about 2.6$\rho_0$
to about 0.8$\rho_0$,
the kaon flow indeed probes the behavior of kaons in a dense nuclear matter.

The magnitude of kaon flow parameter depends also
on the kaon-nucleon cross section.  Increasing the
cross section by a factor of two affects the kaon flow parameter
by about 15 MeV for all cases of the kaon potential.
This effect is comparable to the effect due to the difference
between the kaon potential obtained from the impulse approximation using
the scattering length and from the scalar and vector potentials derived from
the mean-field approximation to the chiral Lagrangian. However, it is still
much smaller than the difference between the cases with and without the kaon
potential and the difference between the cases with and without the
scalar potential. Thus the study of kaon flow is expected to provide useful
information on the kaon potential in a dense matter.

In summary, using the relativistic transport model we have studied the kaon
flow and the associated flow parameter in heavy-ion collisions at SIS/GSI
energies. We have found that the kaon flow is essentially determined during the
expansion stage of heavy-ion collisions. With different kaon potentials as
predicted by current theoretical models, the pattern of kaon flow and the
magnitude of the flow parameter have been investigated. Our results indicate
that the kaon flow is sensitive to the kaon potential.
It is thus very useful in
the future to carry out the flow analysis of the experimental kaon data from
heavy-ion collisions in order to extract information on the kaon potential
in a nuclear medium.

\medskip
We are grateful to Wolfgang Bauer for a critical reading of the manuscript
and helpful suggestions.
This work was supported in part by the National Science Foundation  Grant
No. PHY-9212209 and the Welch Foundation Grant No. A-1110.

\pagebreak

\centerline{\bf Figure Captions}

\begin{description}
\item  {Fig. 1.} Density dependence of the kaon potential in a nuclear
medium.

\item  {Fig. 2.} Average transverse momentum of kaons as a function of
center-of-mass rapidity.

\item  {Fig. 3} Time evolution of the kaon flow parameter. Also shown by the
dashed curve is the central baryon density.
\end{description}

\end{document}